\documentclass[aps,prb,twocolumn,showpacs,floatfix,superscriptaddress]{revtex4-1}

\pdfoutput=1

\usepackage{amsmath}
\usepackage{amssymb}
\usepackage{graphicx}
\usepackage{epstopdf}
\usepackage{latexsym}
\usepackage{subfigure}

\begin{document}

\title{Competing Valence Bond States of Spin-3/2 Fermions on a Strongly Coupled Ladder}
\author{E. Szirmai}
\affiliation{BME-MTA Exotic Quantum Phases Research Group, Institute of Physics, Budapest University of Technology and Economics, 
Budafoki \'ut 8., H-1111 Budapest, Hungary}
\author{H. Nonne}
\affiliation{Quantmetry - Data Science Consulting, 55 rue de la Boétie, 75008 Paris, France}

\begin{abstract}
We study the possible ground state configurations of two strongly coupled chains of charge neutral spin-3/2 fermionic atoms interacting via short range van der Waals interaction. The coupling between the two chains is realized by relatively large hopping amplitude. Exploiting that such a ladder configuration can be mapped to an effective one-band model we analyze the emerging ground states of the system. We show that various spatially inhomogeneous states, valence bond states, plaquette states compete depending on the filling and the ratio of the interaction strengths in the singlet and quintet scattering channel. We find that a Luttinger liquid state is the ground state of the strongly coupled ladder in an extended region of the parameter space, and we also show that a topologically nontrivial charge Haldane state can emerge in the strongly coupled ladder at quarter and three-quarter fillings.
\end{abstract}

\pacs{67.85.-d, 03.75.Ss, 72.15.nj}
\date{\today}
\maketitle

\section{Introduction}

The study of high spin ultracold atomic systems provides powerful means of understanding a series of open questions and fundamental phenomena of strongly correlated systems. One of the most important advantage of ultracold atomic systems
is that they open a possibility to simulate various complicated many particles systems even in model level within well controllable circumstances.~\cite{bloch08a,ketterle08a,lewenstein07a} Accordingly, in principle, a series of nontrivial, exotic magnetic phases~\cite{Marston1989, Harada2003, chen05a, gorshkov10a, wu10c, nascimbene12a} like dimer, trimer, or plaquette orders, resonating valence bond states, or spin liquid states can be studied. This fact triggers not only the experimental research of the high spin systems but also the theoretical studies predicting various expected exotic states of these high spin matter. The interaction between charge neutral atoms in an ultracold atom experiment can be described by weak, short range van der Waals interaction, therefore the atomic cloud can be described by a Hubbard-like model with nearest neighbor hopping and on-site interaction. Accordingly, theoretical studies usually based on various generalization of the Hubbard model to its high spin analogues.

The physics of spin-1/2 particles is the most studied due to its unquestionable importance in condensed matter physics. Spin-3/2 fermionic systems provide the first step towards the physics of fermions with higher spin, and today it has also a rather extended literature (cf. Refs.~\onlinecite{wu06a,tu06a,szirmai11a}). Nevertheless, for two- or three-dimensional systems the applicable methods are mostly limited to mean-field analysis or very demanding numerical methods. Additionally the underlying lattice geometry plays very important role. Therefore our knowledge is rather poor in the higher dimensional cases. Contrary, the one-dimensional (1D) phase diagram of the spin-3/2 Hubbard-like chain is mostly done~\cite{wu03a,lecheminant05a, wu05a, wu06a,capponi07a,nonne10a,nonne11a,barcza12a,nonne13a} due to the powerful methods applicable in 1D, like bosonization, renormalization group as analytical tools, or density matrix renormalization group (DMRG) as numerical tool. 

Between the one- and two-dimensional cases the system of coupled chains, or ladders can exhibit completely new phases, that exist neither in one nor in two dimension. As we will see bellow, this is the case even when the ladder system can be mapped to an effective one-dimensional model.

Assuming Hubbard-like interaction between the fermions, i.e. the Hamiltonian contains only on-site scattering terms,  a ladder configuration can be generated by two chains coupled only via interchain hopping $H_\perp$:
\begin{equation}
\label{eq:ham1}
H = \sum_{j=1,2} \left( H^{(j)} + H_\perp \right) .
\end{equation}
The hopping term along the rungs of the ladder is:
\begin{equation}
H_\perp =  t_\perp \sum_{\sigma ,i} \left( c^{1\dagger}_{\sigma,i} c^2_{\sigma,i} + H.c. \right).
\end{equation}
where $c_{\sigma,i}^{j \dagger}$ ($c_{\sigma,i}^{j}$) creates (annihilates) a spin-3/2 fermion on site $R_i$ of the chain $j$ ($j=1$ or 2), with spin $z$ component $\sigma$. Because of the on-site interaction, the two-particle wave function of the scattering particles has a symmetric spatial part, therefore it has to be antisymmetric in its spin part. Accordingly, the scattering processes take place in two spin channels, the total spin singlet and the total spin quintet channels. If $S$ denotes the total spin of the two colliding particles, $P_{S}^{(j)}$ projects to the total spin-S subspace of the chain $j$ and can be expressed with the pairing operators: $P_S^{(j)}=\sum_{m,i} P_{Sm,i}^{(j)\dagger} P_{Sm,i}^{(j)}$. $P_{Sm,i}^{(j)\dagger} = \sum_{\sigma,\sigma'} \left< \frac32, \frac32; \sigma, \sigma' |S,m\right> 
c_{\sigma,i}^{j\dagger} c_{\sigma',i}^{j\dagger}$, where $\left< \frac32, \frac32; \sigma, \sigma' |S,m\right> $ are the corresponding Clebsch-Gordan coefficients. 
The corresponding intrachain Hamiltonian has the form
\begin{equation}
H^{(j)} = -t \sum_{\sigma ,i} \left( c^{j\dagger}_{\sigma,i+1} c^j_{\sigma,i} + H.c. \right) + g_0 P_{0}^{(j)} + g_2 P_{2}^{(j)},
\end{equation}
where $g_0$ and $g_2$ characterize the strength of the interaction in the singlet, and quintet scattering channels, respectively.

The Hamiltonian \eqref{eq:ham1} conserves the total particle number, therefore it has a U(1) symmetry. It becomes a local gauge symmetry if $t_\perp = t = 0$, reflecting that the on-site interaction terms conserve the particle number on each site independently. Furthermore, for zero interchain hopping ($t_\perp = 0$, $t\neq 0$) the particle number is conserved on both legs independently leading to a U(1)$^1 \times$U(1)$^2$ symmetry, where the superscript 1 and 2, respectively, labels the two legs of the ladder. The system exhibit a $Z_2$ symmetry according to the equivalence of the two legs. In the spin space the symmetry of the Hamiltonian --- even for general couplings $g_0$, and $g_2$ --- is higher than an obvious SU(2) spin rotational invariance, it has an SO(5) symmetry.~\cite{lecheminant05a, wu05a, wu06a} Depending on the filling --- especially on bipartite lattices --- further symmetries can emerge like SO(6), SO(7), SO(8),~\cite{lecheminant05a, wu05a, wu06a} and most importantly in the special case of equal couplings $g_0=g_2$ the Hamiltonian \eqref{eq:ham1} has an SU(4) symmetry. In Ref.~\onlinecite{wu06a}. C. Wu has shown that in case of the special coupling ratio $g_0=5 g_2$ a hidden symmetry emerges in the half filled system: an SU(2) symmetry in the charge sector. This SU(2) symmetry is analogous to the one described first by Yang in Ref.~\onlinecite{yang89a}. (see also Ref.~\onlinecite{hubbard_model_book_1}.) for the half-filled two-component Hubbard model. For spin-3/2 system the hidden SU(2) symmetry in the charge sector corresponds to the degeneracy of the single site empty, doubly occupied and fully occupied states without hopping and chemical potential terms. The chemical potential term plays similar role in the charge sector than an external magnetic field in the spin sector, and breaks down the charge SU(2) symmetry to U(1). When the deviation from the special ratio of the couplings $g_0=5g_2$ is sufficiently small, a charge Haldane state can be identified.~\cite{nonne10a,nonne11a}

Large interchain hopping amplitude can drive the system to the state in which the chains are strongly coupled and can be described by an effective one-band model. However, the nature of the quasi-particles of the effective model are depend on the filling of the particles. In this paper we derive and exploit this mapping, and analyze the possible and competing states of the spin-3/2 fermionic ladder based on the results for the 1D chain. We show that numerous unusual site and bond orders can stabilize along the ladder, and among them one can find such exotic states as wattle or alternating dimer/trimer/quatrimer order, various plaquette states, just as superfluid or molecular density waves with BCS-like pairs, or 4-particle quartets. We show that switching on the tunneling between the two chains can effect the physical properties  fundamental level, such as insulator uncoupled chains can become metallic in presence of interchain coupling. 
We also show that such characteristic phases of the one-dimensional chain like the Luttinger liquid phase just as the charge Haldane state are robust, and remain stable in the presence of strong interchain hopping. Nevertheless, these phases are not identical to their single chain counterparts, according to the quasi-particles which characterize them and which mix the two chains. 

It is important to emphasize, that our mapping is valid when the interchain hopping is comparable with or even
(significantly) smaller than the intrachain hopping.  However, the condition depends on the
filling, even the full lower band case --- which requires the strongest interchain hopping --- can be realized with interchain coupling comparable
with intrachain hopping.

\section{Effective chain model}

In this section we show that the two-leg ladder problem can be described by an effective chain model for sufficiently large rung hopping. To this end let us start with two uncoupled chains: $t_\perp = 0$. We use the Fourier transform $c_{\sigma,i}^j = \frac{1}{\sqrt{\mathcal{N}_s}} \sum_{k\in 1BZ} c_{\sigma,k}^j e^{\mathrm{i}kR_i a_0}$, where $k$ belongs to the 1st Brillouin zone, $\mathcal{N}_s$ is the number of the sites along a chain with length $L=\mathcal{N}_s a_0$, and $a_0$ is the lattice constant. In case of 4-component fermions the filling of the ladder is $f_l=N_f/8\mathcal{N}_s$, where $N_f$ denotes the number of the fermions. Now the kinetic term of the independent chains is diagonal:
\begin{equation}
 H_0^{(j)} = \sum_{\sigma ,k} \epsilon_k c_{\sigma,k}^{j\dagger} c_{\sigma,k}^j.
\end{equation}
Due to the $Z_2$ symmetry $\epsilon_k=\epsilon_k^{(1)}=\epsilon_k^{(2)}$, and similarly the Fermi momentum is  $k_F=k_F^{(1)}=k_F^{(2)}=\pi f_l /a_0$.

\begin{figure}[t]
\centering
 \includegraphics[scale=0.3]{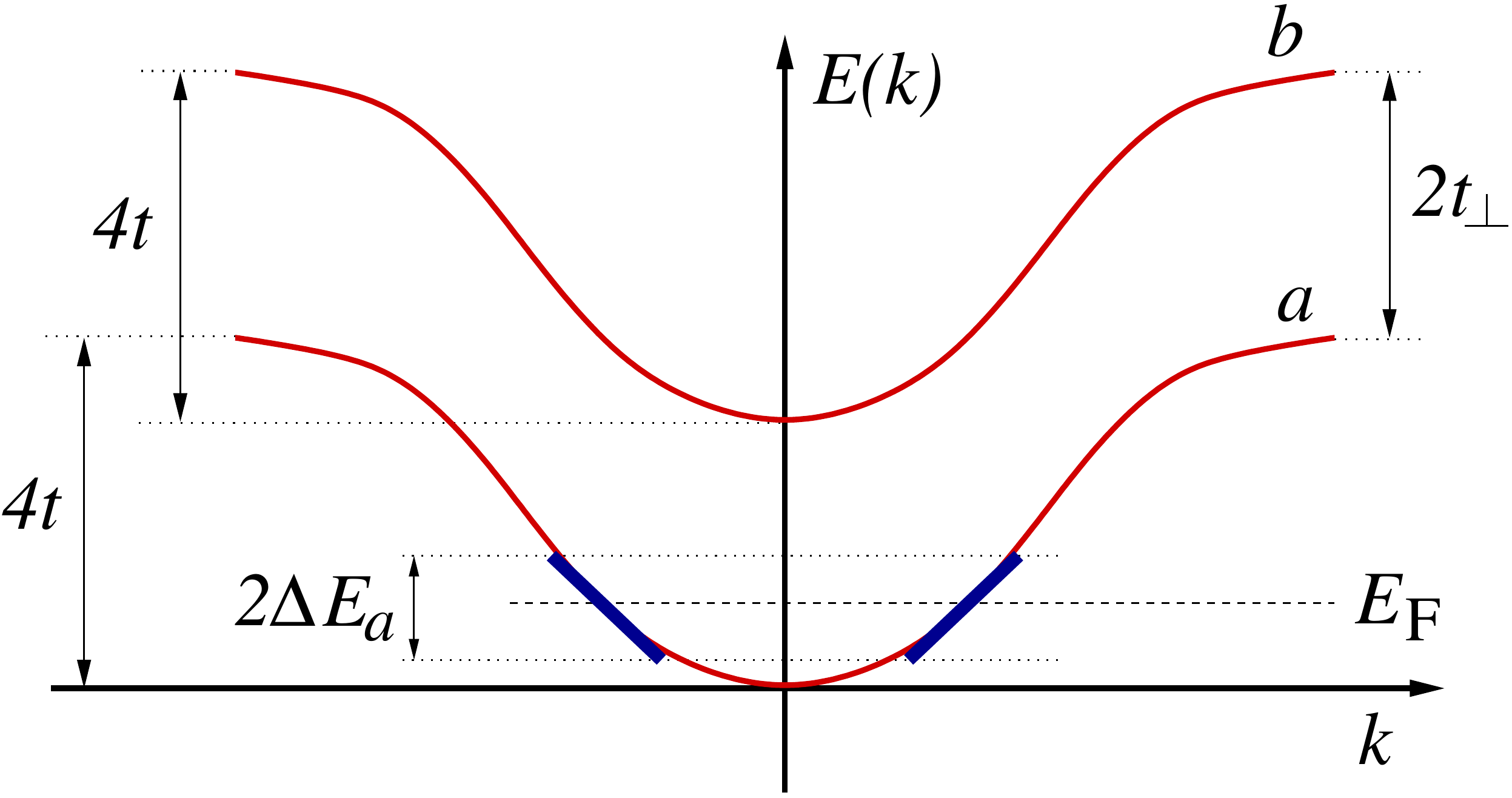}
\caption{(Color online) The spectrum of a strongly coupled ladder, for $f_l<1/2$.}
\label{fig:spectra}
\end{figure}

If $t_\perp \neq 0 $, the noninteracting terms can be diagonalized in the bonding-antibonding basis by introducing the following new operators:
\begin{eqnarray*}
 a_{\sigma,k} & = & (c_{\sigma,k}^1 - c_{\sigma,k}^2)/\sqrt{2}, \\
 b_{\sigma,k} & = & (c_{\sigma,k}^1 + c_{\sigma,k}^2)/\sqrt{2}.
\end{eqnarray*}
The noninteracting part of the Hamiltonian becomes 
\begin{equation}
 H_0 = \sum_{\sigma,k} \left[ (\epsilon_k - t_\perp) a_{\sigma,k}^\dagger a_{\sigma,k} + (\epsilon_k + t_\perp) b_{\sigma,k}^\dagger b_{\sigma,k} \right] .
\end{equation}
Within a tight binding approximation the $\epsilon_k$ spectrum can be written as $\epsilon_k = -2 t \cos (k a_0)$. The degeneracy of the spectrum is resolved due to the interchain hopping, the two cosine curves are shifted by $\pm t_\perp$, respectively, as it is shown in Fig.~\ref{fig:spectra}. 
Now, depending on the tunneling amplitudes $t$ and $t_\perp$, and the filling $f_l$, different cases can be distinguished. 

It is possible to tune the hoppings and the filling to $2t_\perp > E_F$ in order to realize the case when there is no particle in the upper band. For simplicity, here and in the following the energy is measured from the bottom of the lower band. It is possible only for $f_l\leq 1/2$ fillings. In this case there are only two Fermi points $\pm k_a$. We would like to determine the low energy properties that characterize the system, accordingly, it is sufficient to take into account the states in the vicinity of the Fermi energy, and the spectrum can be linearized around the new Fermi points with the bandwidth $2\Delta E_a$. The bandwidth $\Delta E_a$ is restricted only by the validation of the linearization of the cosine curve close to the Fermi points. Now, if $2t_\perp > E_F + \Delta E_a$, the scattering processes between the two bands can be neglected, and one arrive to an effective one-band model. This situation is illustrated in Fig.~\ref{fig:spectra}, and the effective Hamiltonian is given below in Eq.~\eqref{eq:ham-1a}

Decreasing $t_\perp$ or increasing the filling $f_l$, one can reach a region where the linearization is still valid in the lower band, but the role of the states from the upper band can not be neglected any more. If $E_F-\Delta E_a \leq 2t_\perp < E_F + \Delta E_a$, linearization can not be used for the upper band, its quadratic curvature has to be treated.~\cite{balents96a, lin98a} Further decreasing $t_\perp$ the filling of the upper band is increasing, two new Fermi points occur $\pm k_b$, and for sufficiently large band filling the linearization becomes valid around these Fermi points, too, with band width cut-off $\Delta E_b$. Accordingly, in this regime both bands are partially filled, and the Fermi points are $k_a= \pi f_a/a_0$, $k_b= \pi f_b /a_0$, where $f_a$, $f_b$ denotes the filling of the band $a$ and $b$, respectively, and they are determined by the rung hopping $t_\perp$ and the filling of the ladder $f_l$.~\cite{szirmai06a, balents96a, lin98a} In these cases the one-band picture breaks down.

If the filling $f_l > 1/2$, and the tunneling between the chains is $2t_\perp \geq 4t$, the lower band is completely filled and there are two Fermi points in the upper band $\pm k_b$ determined by the filling $f_l$. When the linearization around these Fermi points is reliable, the lower band does not give contribution to the low energy physics. Therefore the effective model of the system is again a one-band model. The effective Hamiltonian is exactly the same as it was in the case of the empty upper band. However, the Hilbert space is different in the two cases. Since in the former (empty upper band) case the effective Hamiltonian acts on the bonding states, in the latter (full lower band) case it acts on the antibonding states, and the emerging ladder states exhibit different properties.  If  $2t_\perp < 4t$, for $4t < E_F - \Delta E_b$ the processes with the states of the lower band still do not give contribution, contrary, for $E_F +\Delta E_b > 4t > E_F - \Delta E_b$ one has to take into account the scattering processes between the two bands even within the low energy approach, and has to treat the quadratic curvature of the lower band.

We emphasize, that the above mapping does not require extremely strong interchain tunneling. The ladder is strongly coupled in the sense that the coupling between the two legs has to be sufficiently large to drive the system into the state where the interband interactions can be neglected. We have seen above, that this situation can be realized with interchain coupling comparable
with intrachain hopping not only in the empty upper band case but even in the full lower band case.

\begin{figure}[t]
\centering
\includegraphics[scale=0.24]{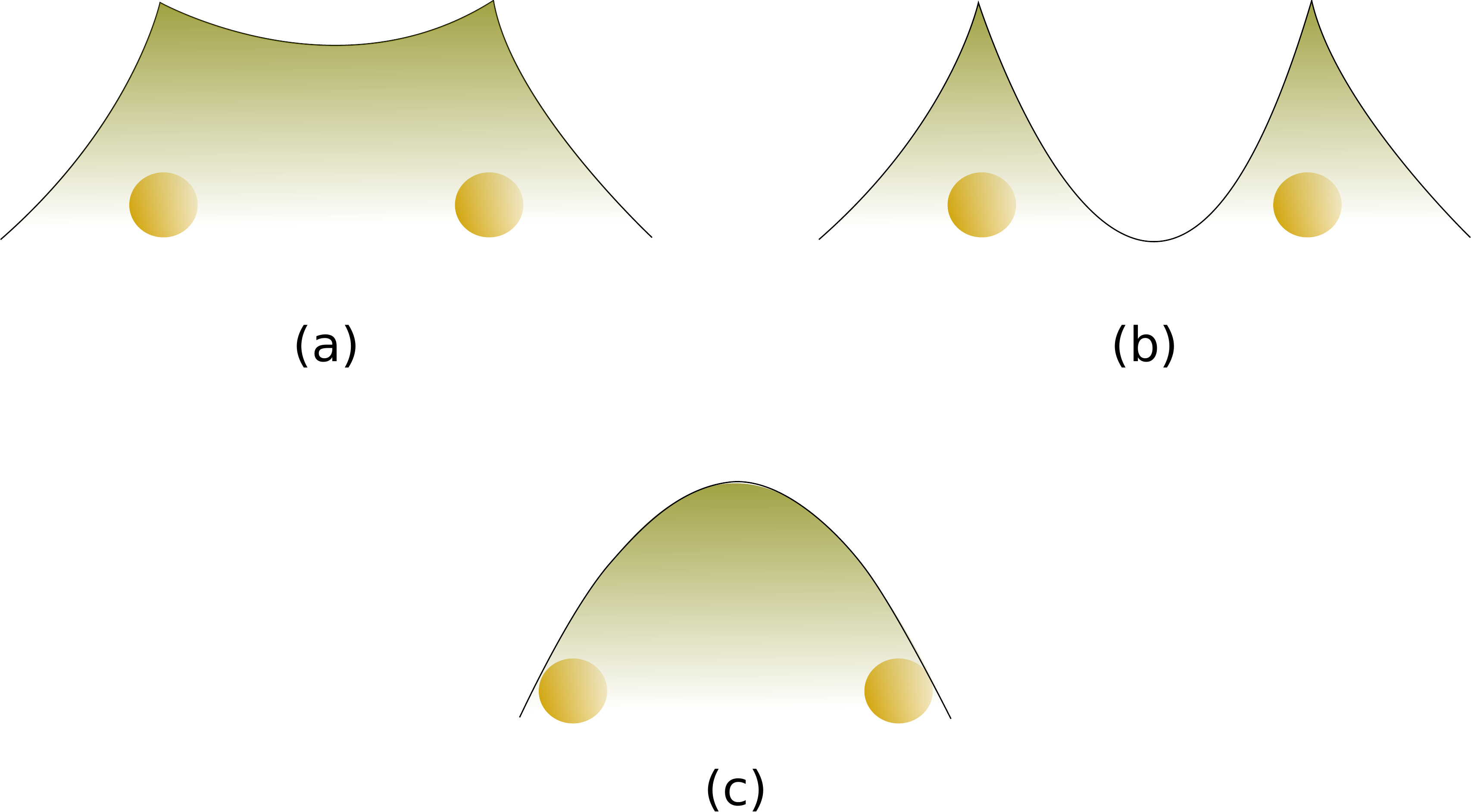}
\caption{(Color online) Schematic density profile of the quasi-particles in the bonding (a) and anti-bonding (b) states along the rungs. Similar density profile characterizes the quasi-particle density in the various density wave like states, namely in a site-centered (b) or a bond-centered (c) density wave, or the coexistence of the two (a).}
\label{fig:densities}
\end{figure}

\section{Possible phases of the strongly coupled ladder} 

In the following we will give a detailed discussion of the possible states of the strongly coupled ladder, when either the  upper or the lower band is empty and does not give any contribution, therefore an effective one-band model describes the system. 
The one-band spin-3/2 system was studied in details in Refs.~\cite{lecheminant05a,wu05a,wu06a} The emerging phases in the ladder configuration can be analyzed in view of the one-band orders, with the mapping the corresponding order parameters to the ladder states that can lead to novel, exotic states. The analysis of the weakly coupled chains --- i.e. the intermediate states which occur as a consequence of the scatterings between the bonding and antibonding states --- are beyond the subject of the present paper, they will be studied elsewhere.

The SU(4) symmetric case, when $g_0=g_2$, has special interest because of their importance in the experiments with alkaline earth atoms. The electronic spin of alkaline earth atoms is zero, therefore, their hyperfine spin is simply determined by the nuclear spin. The dominant interaction between these atoms is a van der Waals interaction generated by induced electronic dipole moment of the atoms, which does not depend on the hyperfine spin. The spin independent scattering processes lead to an SU(4) symmetric interaction between the alkaline earth atoms. Accordingly, in the following, we will discuss the SU(4) symmetric case separately. 

\subsection{Empty upper band}

Let us consider first the case when there is no particle in the upper band and the scattering processes between the two bands can be neglected. We have seen, that it is possible for $2t_\perp > E_F + \Delta E_a$ and for $f_l\leq 1/2$. Due to the missing bonding ($b$) states, $f_a=2f_l$, and the Fermi energy $E_F$ is determined by only the hopping $t$ and the filling: $E_F=-2t\mathrm{cos}(\pi f_a)=-2t\mathrm{cos}(2\pi f_l)$. In this case the only nonzero terms of the Hamiltonian in the bonding-antibonding basis are those that consist only $a$ operators. Therefore the effective one-band Hamiltonian is
\begin{multline}
\label{eq:ham-1a}
 H = -t \sum_{\sigma,i} \left( a_{\sigma,i}^\dagger a_{\sigma,i+1} + H.c. \right) - t_\perp \sum_{\sigma,i} n_{\sigma,i}^a \\ + \frac{g_0}{2} P_0^a + \frac{g_2}{2} P_2^a ,
\end{multline}
where $P_S^a=\sum_{m,i} P_{Sm,i}^{a\dagger} P_{Sm,i}^{a\phantom\dagger}$, and $P_{Sm,i}^{a\dagger} = \sum_{\sigma,\sigma'} \left< \frac32, \frac32; \sigma, \sigma' |S,m\right> 
a_{\sigma,i}^\dagger a_{\sigma',i}^\dagger$.
This Hamiltonian acts only on the states of the lower band $a$, and with this restricted Hilbert space it describes a one-dimensional, 4-component Hubbard model. 
Without intrachain hopping $t$, the Hamiltonian in Eq.~\eqref{eq:ham-1a} is diagonal in real space and it is reduced to a one-site problem in the bonding-antibonding basis. Its eigenstates and the corresponding eigenenergies are summarized in Tab.~\ref{tab:1site-problem}. 

The emerging states strongly depend on filling, therefore we discuss the possible phases for incommensurate fillings and for various characteristic commensurate fillings.

\begin{table}
\centering
\begin{tabular}{|c|c|}
\hline
$\Psi_i$ & $E_i$ \\
\hline
\hline
$| 0 \big>$ & $E_0=0$ \\
\hline
$a_{\sigma,i}^\dagger | 0 \big>$ & $E_1 = - t_\perp$ \\
\hline
$ P_{00,i}^{(a)\dagger} |0 \big>$ & $E_3= -2 t_\perp + g_0$ \\
\hline
$ P_{2m,i}^{(a)\dagger}|0 \big>$, $m=0$, $\pm 1$, $\pm 2$ & $E_2= -2 t_\perp + g_2$ \\
\hline
$f_{\sigma,i}^{(a)\dagger}| 0 \big>$ & $E_4= -3 t_\perp + \frac12 g_0 + \frac52 g_2$ \\
\hline
$a_{\frac32,i}^\dagger a_{\frac12,i}^\dagger a_{-\frac12,i}^\dagger a_{-\frac32,i}^\dagger | 0 \big>$ & $E_5 = -4t_\perp + g_0 + 5 g_2$ \\
\hline
\end{tabular}
\caption{$\Psi_i$'s are the eigenstates of the one-site problem and $E_i$ are the corresponding eigenenergies. In line 4. $f_{\sigma,i}^{(a)\dagger}| 0 \big> =a_{\sigma,i} a_{\frac32,i}^\dagger a_{\frac12,i}^\dagger a_{-\frac12,i}^\dagger a_{-\frac32,i}^\dagger | 0 \big>$.
}
\label{tab:1site-problem}
\end{table}

\subsubsection{Quarter filled ladder} 

\begin{figure}[t]
\centering
\includegraphics[scale=0.28]{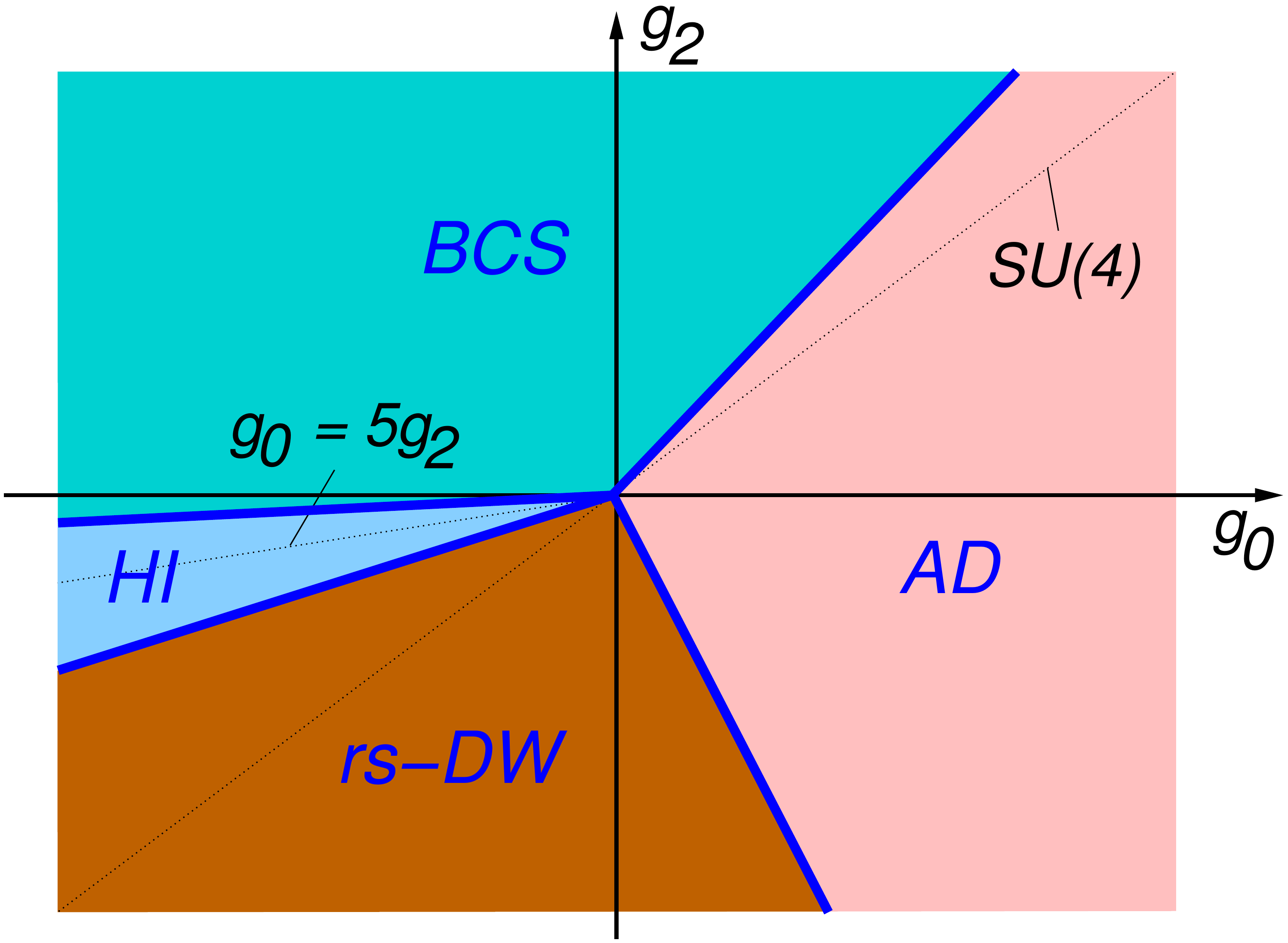}
\caption{(Color online) The phase diagram of the strongly coupled ladder at $f_l=1/4$ filling. For the notations see the text.}
\label{fig:phase-diag1}
\end{figure}

When the ladder is quarter filled, there is one particle per site on average. In this case the lower band is half-filled, therefore one has to consider the one-band Hubbard model in Eq.~\eqref{eq:ham-1a} at half-filling. The requirement for the hoppings becomes: $2t_\perp > 2t + \Delta E_a$. The phase diagram of the quarter filled system is shown in Fig.~\ref{fig:phase-diag1}.

\begin{figure}[t]
\centering
\subfigure[\, AD state, quarter filled ladder.]{
\includegraphics[scale=0.3]{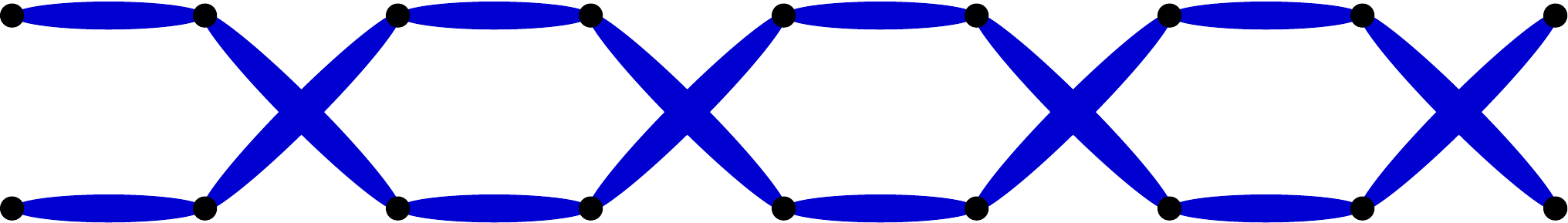}
\label{fig:subfig1}
}
\hskip 2cm
\subfigure[\, rs-DW, quarter filled ladder.]{
\includegraphics[scale=0.3]{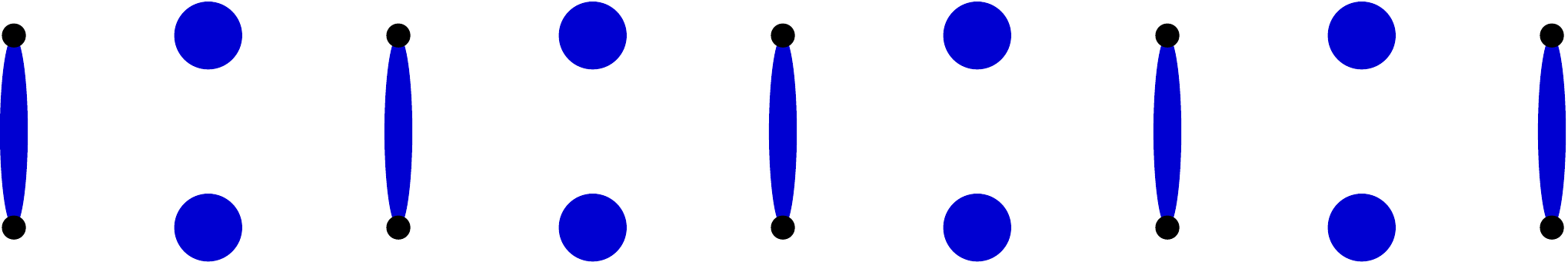}
\label{fig:subfig2}
}
\caption{(Color online) Illustration of the various density wave states at quarter filling. Blue dots at the sites and strings along the bonds denote large particle density at the site and the bond, respectively.}
\label{fig:subfiguresExample1}
\end{figure}

{\it Along the SU(4) line} --- Let us denote $g_0=g_2 =: U/2$. If $U>0$, a spin-Peierls phase stabilizes in the one band model. This is a bond centered atomic density wave (BDW) with $2k_a$ wave vector, and the corresponding order parameter is $\mathcal{O}_{BDW}^a =  e^{\mathrm{i} 2 k_a R_i} \sum_\sigma  \left( a_{\sigma,i}^\dagger a_{\sigma,i+1} + H.c. \right)$. In our original ladder basis this state relates to a $4k_F$-BDW:
\begin{multline}
\label{eq:ad-op}
 \mathcal{O}_{BDW}^a = e^{\mathrm{i} 4 k_F R_i} \sum_\sigma \frac12  \Big( c^{1\dagger}_{\sigma,i+1} c^1_{\sigma,i} + c^{2\dagger}_{\sigma,i+1} c^2_{\sigma,i} \\ - c^{1\dagger}_{\sigma,i+1} c^2_{\sigma,i} - c^{2\dagger}_{\sigma,i+1} c^1_{\sigma,i}  + H.c. \Big),
\end{multline}
where we use the notation $k_F=\frac{\pi}{a_0} f_l$. Since at quarter filling  $e^{\mathrm{i} 4 k_F R_i}=(-1)^i$, in this phase a bond centered and a cross centered density wave of the atoms coexist with $2a_0$ periodicity. Due to the $\pi$ phase difference between the two BDWs, their maximum alternate as shown in Fig.~\ref{fig:subfig1}. Therefore we call this state alternating dimer (AD) phase. This phase is expected to be a nonhomogeneous metallic state contrary to its single chain counterpart which is an insulating state. The difference is a direct consequence of the antibonding nature of the quasi-particles.

If $U<0$, site centered atomic density wave (DW) with $2k_a$ wave vector appears in the one-band model described by the order parameter: $\mathcal{O}_{DW}^a = e^{\mathrm{i} 2 k_a R_i} \sum_\sigma  \left( a_{\sigma,i}^\dagger a_{\sigma,i} \right)$. In the ladder basis this is a $4k_F$-DW:
\begin{equation}
\label{eq:adw-op}
  \mathcal{O}_{DW}^a= e^{\mathrm{i} 4 k_F R_i} \sum_\sigma \frac12  \left( n^1_{\sigma,i} + n^2_{\sigma,i} - c^{1\dagger}_{\sigma,i} c^2_{\sigma,i} - c^{2\dagger}_{\sigma,i} c^1_{\sigma,i} \right) .
\end{equation}
The order parameter of both the rung and the site centered density wave have nonzero expectation value leading to the coexistence of these two types of atomic density wave with $2a_0$ periodicity but with opposite phase. Due to the phase shift of the two density waves, the particle density shows a minimum or maximum, respectively, at the center of every second rungs or at the two sites at the ends of the rungs. This rung-site density-wave (rs-DW) order is illustrated in Fig.~\ref{fig:subfig2} referring to the alternating maximum of the density between the rungs and sites.

{\it For $g_0\neq g_2$} ---Two different phases have been found in this case for repulsive interactions. For dominating $g_0$ the system shows an instability against the formation of the AD phase described by the order parameter Eq.~(\ref{eq:ad-op}), i.e.  the ground state of the SU(4) system is stable in an extended region (see Fig.~\ref{fig:phase-diag1}). For dominating $g_2$ site singlets are formed in the antibonding band described by the order parameter $\mathcal{O}_{BCS}^a=P_{00,i}^{a\dagger}$. This leads to the coexistence of rung and site singlets along the ladder:
\begin{multline}
\label{eq:bcs-op}
 \mathcal{O}_{BCS}^a=  \frac12  \big( c^{1\dagger}_{\frac32,i} c^{1\dagger}_{-\frac32,i} - c^{1\dagger}_{\frac12,i} c^{1\dagger}_{-\frac12,i}  + 
c^{2\dagger}_{\frac32,i} c^{2\dagger}_{-\frac32,i} - c^{2\dagger}_{\frac12,i} c^{2\dagger}_{-\frac12,i}
\\ - c^{1\dagger}_{\frac32,i} c^{2\dagger}_{-\frac32,i} + c^{1\dagger}_{\frac12,i} c^{2\dagger}_{-\frac12,i} - c^{2\dagger}_{\frac32,i} c^{1\dagger}_{-\frac32,i} + c^{2\dagger}_{\frac12,i} c^{1\dagger}_{-\frac12,i}  \big). 
\end{multline}
Accordingly, the singlet pairs are suited along the rungs, however, the pair density is small at the center of the rungs as a consequence of that the pairs consist of two anti-bonding particles. The density profile of the pairs looks like in Fig.~\ref{fig:densities}(b). This state conserves the (discrete) translation invariance of the Hamiltonian, so that can be considered as a superfluid phase. This phase remains the ground state everywhere in the quarter plane of $g_0<0$, and $g_2>0$.

The rs-DW state of the SU(4) symmetric system with attractive interaction is the ground state in an extended region, too. While the AD order collapses directly to the rs-DW order as the singlet coupling $g_0$ is decreasing for fixed negative value of $g_2$, between the BCS and the rs-DW state a Haldane insulating state emerges with nontrivial topology. This state can not be characterized by a local order parameter, instead by a nonlocal string order parameter along the ladder:
\begin{equation}
\label{eq:string_op}
\mathcal{O}_{HI}^a = \mathrm{lim}_{|i-j|\rightarrow \infty} \left< \tilde{n}_i^a \mathrm{e}^{\mathrm{i} \pi \sum_{i<k<j} \tilde{n}_k^a} \tilde{n}_j^a \right> ,
\end{equation} 
where $\tilde{n}_i^a = (n_i^a - 2)/2$. 

The Haldane insulating state appears around the high symmetric line $g_0=5 g_2$, that corresponds to a hidden, extended charge SU(2) symmetry of a half-filled chain, described in Refs.~\onlinecite{wu03a,wu06a,nonne10a} (see also Ref.~\onlinecite{szirmai11a}). The generators of the corresponding SU(2) can be expressed with the singlet pairing operator and the particle number operator as $\mathcal{S}_i^\dagger = P_{00,i}^{a\dagger}$, and $\mathcal{S}_i^z = (n_i^a - 2)/2$, respectively. Therefore $\mathcal{S}_i^z$ is identical with $\tilde{n}_i^a$ in Eq.~\eqref{eq:string_op}. Accordingly, in this state --- similarly to the BCS state ---  singlet pairs are suited along the rungs, however, despite the preserved (discrete) translation invariance along the length of the ladder, there is no quasi-long range pair-order. It is difficult to measure a nonlocal string order parameter, but fortunately, the difference between the Haldane state and the BCS state can catch also through the nontrivial topology exhibited in the former state. The topological nature of the Haldane state can be characterized by the edge pair states occurring at the two end-rungs. These edge states do not carry spin, they are in singlet state, and are formed by antibonding particles, therefore, the pair density shows a minimum at the center of the end rungs.

\subsubsection{General fillings}

Now let us consider the case of general fillings. Commensurability usually comes with emergence of (marginally) relevant umklapp processes that can lead to occurrence of various order with spatial periodicity. Contrary, for incommensurate fillings when the ratio of the number of fermions and the number of sites is not a rational number there are no umklapp processes to take into account.  Above we considered in details the case of the quarter filled ladder and in the following we will see further examples for various commensurate fillings, besides the incommensurate filled system. We consider only the commensurate fillings with $f_l=p/2q$, where $p$ and $q$ are integer, i.e. the effective one-band model is $f_a=p/q$ filled.

\begin{figure}[t]
\centering
\subfigure[\, AT state, 1/6-filled ladder.]{
\includegraphics[scale=0.3]{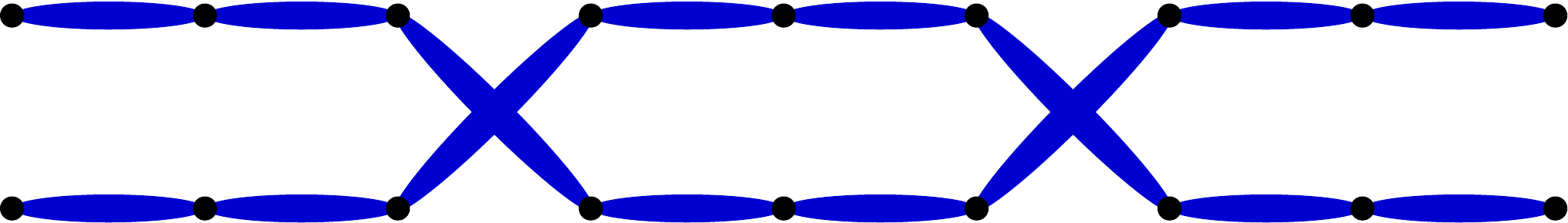}
\label{fig:subfig3}
}
\hskip 2cm
\subfigure[\, 8$k_F$-MDW, 1/8-filled ladder.]{
\includegraphics[scale=0.3]{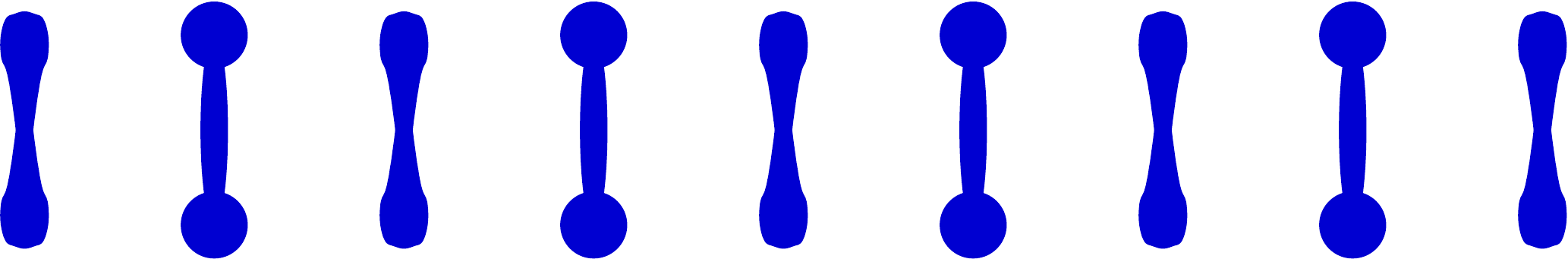}
\label{fig:subfig4}
}
\\
\subfigure[\, MVBS state, $f_l=1/8$.]{
\includegraphics[scale=0.3]{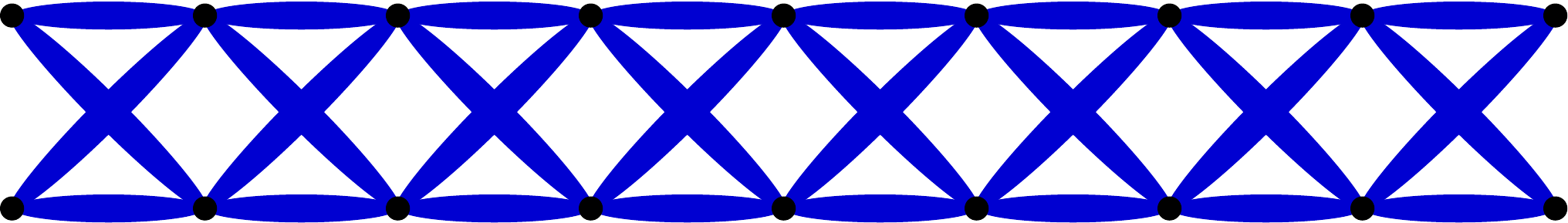}
\label{fig:subfig5}
}
\hskip 2cm
\subfigure[\, AQ state, 1/8-filled ladder.]{
\includegraphics[scale=0.3]{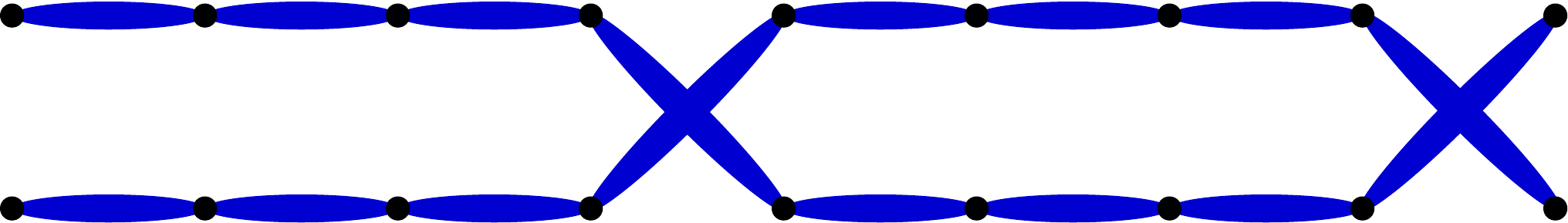}
\label{fig:subfig6}
}
\label{fig:subfiguresExample2}
\caption{(Color online) The same as in Fig.~\ref{fig:subfiguresExample1} for various commensurate fillings.}
\end{figure}

{\it Along the SU(4) line} ---($g_0=g_2 =: U/2$) For repulsive interaction ($U>0$), and for incommensurate fillings the system is a Luttinger liquid. Because of the empty upper band the band degree of freedom is frozen out leading to a 4-component Luttinger liquid, however, the freely fluctuating density fields consist of particles on both legs. For commensurate $f_l=p/2q$ fillings  the system remains homogeneous if $q$ is not less than the number of the spin components $2F+1$ of the fermions.~\cite{szirmai08a} For $F=3/2$ fermions $2F+1=4$, so for $q\geq 4$ no inhomogeneity occurs. Contrary, relevant umklapp processes lead to the occurrence of spatially inhomogeneity for $q< 4$.
In case of empty upper band $f_l\leq 1/2$. For $q=1$, the only possible filling is $f_l=1/2$. For strongly coupled half filled ladder the upper band is empty, and the lower band is completely filled. This case will be discussed at the beginning of the next Section. If $q=2$, the ladder is quarter filled, this case was discussed above. The last remaining case, when spatial inhomogeneity is expected to emerge in the ground state, is if $q=3$, $f_l=1/6$. In this case a phase with three-fold spatial periodicity stabilizes. This phase is characterized by the $4k_F$-BDW order parameter defined in Eq.~(\ref{eq:ad-op}), but now $e^{\mathrm{i} 4 k_F R_i}=e^{\mathrm{i}2\pi/3a_0}$. While a density wave with $4k_F$ wave vector corresponds to a dimerized state for quarter filled system, it leads to an alternating trimerized (AT) order at $f_l=1/6$ filling as shown in Fig.~\ref{fig:subfig3}.~\cite{szirmai08a} This phase --- similarly to the AD state --- is an inhomogeneous metallic state, however, its single chain counterpart is an insulator. 

For attractive interaction ($U<0$) at incommensurate fillings 
two phases compete: SU(4) singlet quartets and a density wave order. 
The order parameter of the quartets in the one-band model is $\mathcal{O}_\mathrm{quart}= a_{\frac32,i}^\dagger a_{\frac12,i}^\dagger a_{-\frac12,i}^\dagger a_{-\frac32,i}^\dagger$, that reads in the ladder basis as
\begin{equation}
\label{eq:ms-op}
\mathcal{O}_\mathrm{quart}= \frac14 \sum_{j_l} (-1)^{\Sigma j_l}\,\,\, c^{j_1\dagger}_{\frac32,i} c^{j_2\dagger}_{\frac12,i} c^{j_3\dagger}_{-\frac12,i} c^{j_4\dagger}_{-\frac32,i}.
\end{equation}
$j_l=1$ or 2 labels the leg of the ladder and $\Sigma j_l=j_1+j_2+j_3+j_4$. The quartets are formed around the sites and along the rungs, but the state does not violates the translation invariance along the ladder. Therefore this is an SU(4) molecular superfluid (MS) state of the quartets. The DW phase is characterized by nonzero order parameter given in Eq.~(\ref{eq:adw-op}). The phase boundary between the DW and the quartets is determined by the specific value of the Luttinger parameter of the charge degree of freedom of the antibonding quasi particles: $K_c=2$. The MS state is stable for $K_c>2$. The parameter $K_c$ characterizes the interaction within a Luttinger liquid description of hydrodynamical approach, and has been determined for the 4-component one-band model in Ref.~\onlinecite{wu05a}.

Along the SU(4) line we found the same behavior for some commensurate fillings. The $f_l=1/8$ case can be described by a quarter filled chain. In this case the leading umklapp processes are 4-particle scatterings that couple only to the charge mode and they do not lead to new phase.~\cite{szirmai05a,lecheminant05a,wu05a,wu06a} Accordingly, the $f_l=1/8$ filled strongly coupled ladder is expected to behave as in case of incommensurate fillings. For other commensurate fillings the umklapps couple the spin and charge sector, therefore, the relevant umklapp processes can further reduce the number of the degree of freedom which play role in the low energy fluctuations. The density wave order parameter $\mathcal{O}_{DW}^a$ couples only to the charge mode which are not modified by the umklapps in this parameter regime. Contrary, the quartets are coupled also to a spin mode which is pinned by the leading order umklapps. Accordingly, in case of relevant umklapps, due to the pinning of the spin fields, the quartets are frozen out, and only the density wave order remains --- as we have seen in case of $f_l=1/4$.

\begin{figure}[t]
\centering
\includegraphics[scale=0.28]{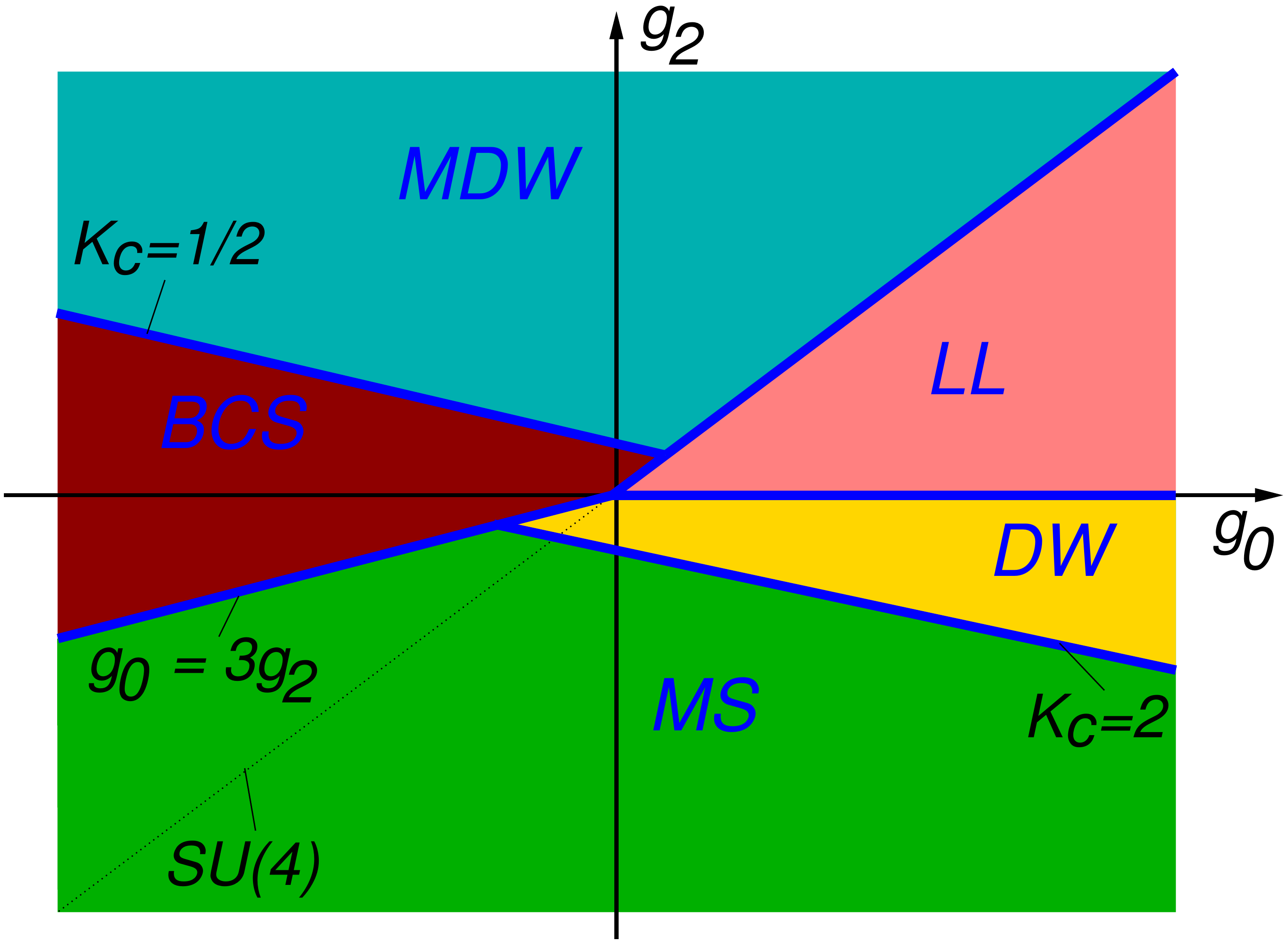}
\caption{(Color online) The phase diagram of the strongly coupled ladder for incommensurate fillings. Note that the value of Luttinger parameter $K_c$ depends on the filling.}
\label{fig:phase-diag2}
\end{figure}

{\it For $g_0\neq g_2$} --- The phase diagram on the $(g_0,g_2)$ plane for general fillings is shown in Fig.~\ref{fig:phase-diag2}. Five different phases appear depending on the values of $g_0$ and $g_2$ for general fillings, and additional two phases emerge at $f_l=1/8$ fillings. 

If $g_0>g_2$ and both couplings are repulsive, the ground state is a Luttinger liquid state. 

If $g_0<g_2$ and $g_0<3g_2$, depending on the value of the Luttinger parameter $K_c$, either a spatially homogeneous superfluid state emerges or density wave stabilizes. For $K_c>1/2$ singlet Cooper pairs form along the rungs and around the sites as we have seen in case of the quarter-filled system. The emergence of these singlets leads to a BCS-like superfluid state. For $K_c<1/2$ site centered (2-atom) molecular density wave (MDW) appears with $4k_a$ wave vector in the one band model described by the order parameter $\mathcal{O}_{MDW}^a= e^{\mathrm{i}4k_a R_i} \sum_{\sigma,\sigma'} a_{\sigma,i}^\dagger a_{\sigma',i}^\dagger a_{\sigma',i} a_{\sigma,i}$. In the ladder basis it has the form:
\begin{equation}
\mathcal{O}_{MDW}^a =  e^{\mathrm{i}8k_F R_i} \frac14 \sum_{j_l} \sum_{\sigma,\sigma'}  (-1)^{\Sigma j_l} \,\,\,c^{j_1\dagger}_{\sigma,i} c^{j_2\dagger}_{\sigma',i} c^{j_3}_{\sigma',i} c^{j_4}_{\sigma,i}.
\end{equation} 
The operator products consisting as many creation operators on a leg as annihilation operators --- e.g. $c^{1\dagger} c^{2\dagger} c^{1} c^{2}$, or $c^{1\dagger} c^{1\dagger} c^{1} c^{1}$ --- describe symmetric rung centered density wave with similar density profile along the rung as in Fig.~\ref{fig:densities}(a). The operator products consisting either more or less creation operators than annihilation operators on the same leg --- e.g. $c^{1\dagger} c^{1\dagger} c^{2} c^{1}$ --- describe skewed rung centered density waves with asymmetric density profile along the rungs. Accordingly, symmetric and skewed rung centered and site centered molecular density wave coexist along the ladder with $8k_F$ wave vector, and the skewed rung MDW has a $\pi$ phase shift. This phase is illustrated in Fig.~\ref{fig:subfig4} for the special $f_l=1/8$ filling, where $8k_F=8\pi/8a_0$ corresponds to a $2a_0$ periodicity. However, at 1/8-filling the lower band is quarter filled, and $8k_a$ umklapp processes become relevant for $K_c<1/2$ in the one band model. When they scale to large repulsive values, they can suppress the original orders and stabilize new ones. Therefore, instead of the $8k_F$-MDW state, bond centered molecular density wave (BMDW) can appear in an extended region of the ($g_0,g_2$) parameter space. The phase boundaries depend on the strength of the umklapp scatterings and $g_0$, $g_2$. The corresponding order parameter:
\begin{multline}
\mathcal{O}_{BMDW}^a =  e^{\mathrm{i}8k_F R_i} \frac18 \sum_{j_l} \sum_{\sigma,\sigma'}  (-1)^{\Sigma j_l} \,\,\, \\ \cdot (c^{j_1\dagger}_{\sigma,i} c^{j_2\dagger}_{\sigma',i+1} c^{j_3}_{\sigma',i} c^{j_4}_{\sigma,i+1} + c^{j_1\dagger}_{\sigma,i+1} c^{j_2\dagger}_{\sigma',i} c^{j_3}_{\sigma',i+1} c^{j_4}_{\sigma,i}).
\end{multline} 
This order parameter describes two independent $8k_F$-BMDW with $\pi$ phase shift which leads to a 2-atom molecular valence bond solid (MVBS) state which preserves the discrete translational invariance of the Hamiltonian along the ladder as illustrated in Fig.~\ref{fig:subfig5}. 

If $g_0>3g_2$ and $g_2<0$, depending on the value of the Luttinger parameter in the charge channel $K_c$, formation of SU(4) singlet quartets leads to a molecular superfluid state (for $K_c>2$) or $4k_F$-DW (for $K_c<2$) characterizes the ground state of the system. These are the same phases that we have seen along the SU(4) line for incommensurate fillings described by the order parameters in Eqs.~(\ref{eq:bcs-op}) and (\ref{eq:adw-op}), respectively. Again, at 1/8-filling the $4k_F$-DW can be suppressed by strong repulsive umklapp processes and instead a $4k_F$-BDW state can emerge with order parameter defined in Eq.~(\ref{eq:ad-op}). However, at $f_l=1/8$ filling $4k_F=4\pi/8 a_0$ corresponds to 4$a_0$ periodicity along the ladder leading to an alternating quatrimer (AQ) state as shown in Fig.~\ref{fig:subfig6}. This state again is a nice example for an inhomogeneous metallic state which has an insulating single chain counterpart. Here we note that the alternating dimer, trimer and quatrimer states are nice examples for the unusual phenomena, when an insulating state of the chain becomes metallic as the interchain tunneling is switched on. This is a consequence of the antibonding coupling between the chains --- realized by quasiparticles in the antibonding state, and we will see, that in case of bonding coupling the insulating characteristic of the single chain is preserved.

\subsection{Full lower band}

When the lower band is completely filled, the anti-bonding degree of freedom is frozen out and the dominant fluctuations can come only from the upper band. Accordingly, the low energy physics is determined by the bonding quasi-particles which can again be described by an effective chain model, the same as Eq.~\eqref{eq:ham-1a} with the exchange of $a$ to $b$. However, one must keep in mind that there is an underlying fully filled anti-bonding band determining the main gap in the system. At half filling this is the only gap in the system, and the emerging order does not depend on the $g_0$, $g_2$ couplings in the weak interaction limit. This order is characterized by four particle quartets along the rungs forming SU(4) singlets. These SU(4) singlet rung-quartets have low quartet density at the center of the rungs. In case of large hopping along the rungs, the rung-SU(4) singlets of the anti-bonding quasi-particles can not be affected by the particles in the bonding states. Instead, the bonding particles create independent orders on top of the robust anti-bonding rung-SU(4) singlets. 

Based on the analysis in the previous section here we analyze the nature of the emerging phases in the case when the contribution of the (lower) anti-bonding band to the low energy physics can be neglected. The quasi-particles in the fully filled anti-bonding band are in the ground state with respect to their Hamiltonian Eq.~\eqref{eq:ham-1a}. The low energy physics is determined by the bonding quasi-particles in the upper band derived by the same Hamiltonian with the exchange of the operators $a$ to $b$.

\subsubsection{3/4-filled ladder}

\begin{figure}[t]
\centering
\includegraphics[scale=0.28]{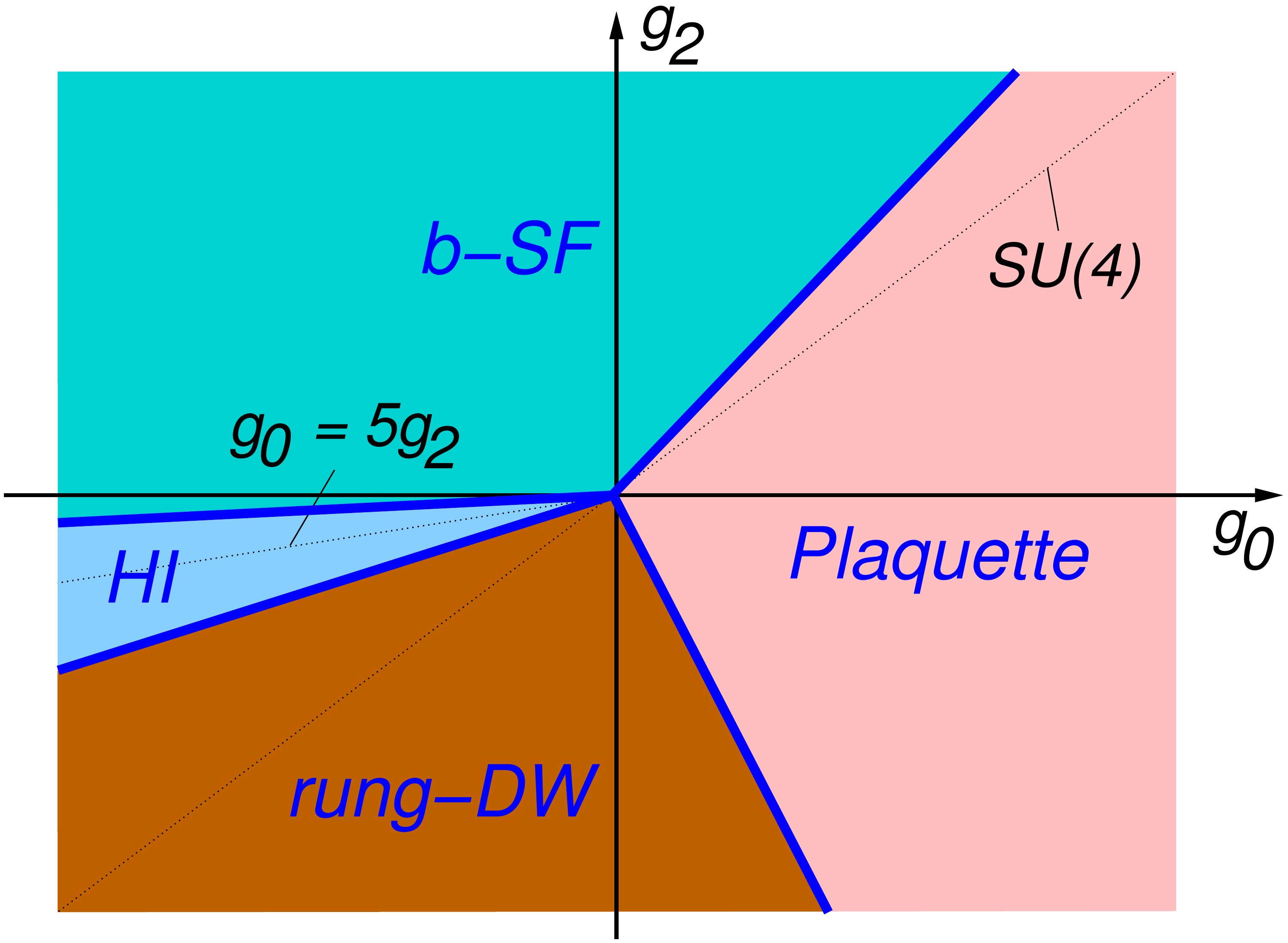}
\caption{(Color online) The phase diagram of the strongly coupled ladder at $f_l=3/4$ filling. For the notations see the text.}
\label{fig:phase-diag3}
\end{figure}

At $f_l=3/4$ filling the structure of the phase diagram is similar to the one of the $f_l=1/4$ filled system. Since the Hamiltonian has the same form, only the Hilbert space is different, the phase boundaries are coincides in the two cases, however, the emerging phases are different. The phase diagram is shown in Fig.~\ref{fig:phase-diag3}.

\begin{figure}[t]
\centering
\subfigure[\, Screwed plaquette state, 3/4-filled ladder.]{
\includegraphics[scale=0.3]{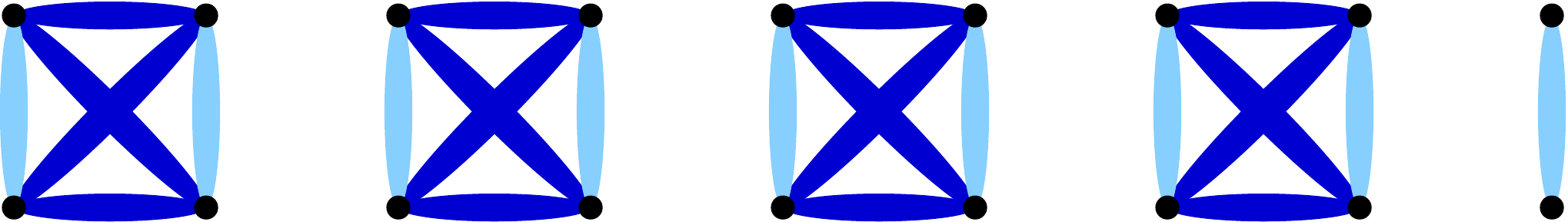}
\label{fig:subfig7}
}
\hskip 2cm
\subfigure[\, Rung-DW, 3/4-filled ladder.]{
\includegraphics[scale=0.3]{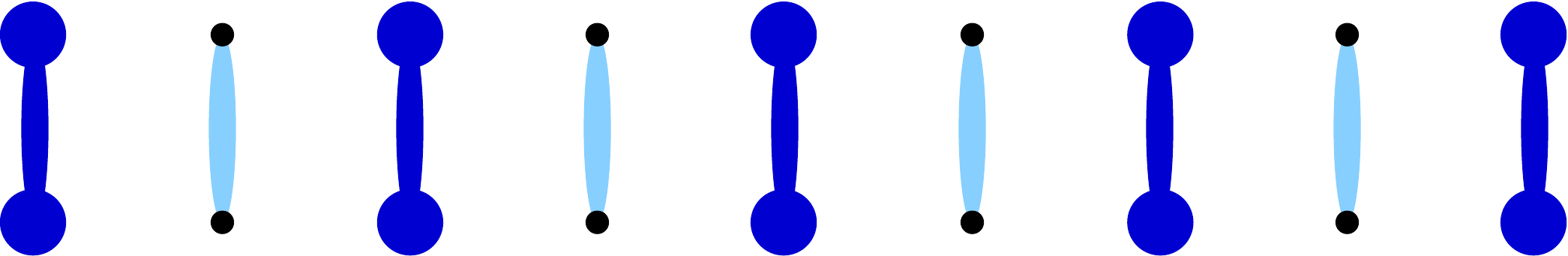}
\label{fig:subfig8}
}
\caption{(Color online) The same as in Fig.~\ref{fig:subfiguresExample1} for $f_l=3/4$. The underlying SU(4) singlet rung-quartets are also indicated with light blue strings.}
\label{fig:subfiguresExample3}
\end{figure}

In the $2k_b$ bond density wave of the bonding quasi-particles there coexist a bond centered and a cross centered density wave without phase shift leading to a screwed plaquette phase --- referred as Plaquette phase on Fig.~\ref{fig:phase-diag3}. This order is illustrated in Fig.~\ref{fig:subfig7}, where the underlying SU(4) rung-quartets are also indicated with light blue. Since  the filling of the upper band is $f_b=2(f_l-1/2)$, $k_b=2k_F-\pi/a_0$.  Accordingly, this is a $4k_F$-BDW, and the corresponding order parameter is:
\begin{multline}
\label{eq:plaquette-op}
 \mathcal{O}_{BDW}^b= e^{\mathrm{i} \left(4 k_F - \frac{2\pi}{a_0}\right) R_i} \sum_\sigma \frac12  \Big( c^{1\dagger}_{\sigma,i+1} c^1_{\sigma,i} + c^{2\dagger}_{\sigma,i+1} c^2_{\sigma,i} \\ + c^{1\dagger}_{\sigma,i+1} c^2_{\sigma,i} + c^{2\dagger}_{\sigma,i+1} c^1_{\sigma,i}  + H.c. \Big).
\end{multline}
Note that $e^{-\mathrm{i} \frac{2\pi}{a_0} R_i}=1$, what will be dropped in the following. We can see, that at f=3/4 filling the insulating character of the single chain is preserved in presence of sufficiently strong intrachain tunneling.

The $2k_b$ density wave state can be characterized by alternating density maximum at every second rungs as in Fig.~\ref{fig:subfig8}.  Along the rungs the density profile follows the one sketched in Fig.~\ref{fig:densities}(a). This phase is a nice example for a pure rung-DW. The corresponding order parameter:
\begin{equation}
\label{eq:rung-dw-op}
  \mathcal{O}_{DW}^b= e^{\mathrm{i} 4 k_F R_i} \sum_\sigma \frac12  \left( n^1_{\sigma,i} + n^2_{\sigma,i} + c^{1\dagger}_{\sigma,i} c^2_{\sigma,i} + c^{2\dagger}_{\sigma,i} c^1_{\sigma,i} \right) .
\end{equation}

The superfluid and the Haldane state are very similar to the ones at quarter filling, however, the pairs --- which play crucial role in both phases --- consist of two bonding quasi-particles instead of two anti-bonding ones. Accordingly, the density profile of the pairs along the rungs looks like in Fig.~\ref{fig:densities}(a). The superfluid phase can simply be characterized by the nonzero order parameter of site and rung-singlet pairs:  
\begin{multline}
\label{eq:sf2-op}
 \mathcal{O}_{SF}^b =  \frac12  \big( c^{1\dagger}_{\frac32,i} c^{1\dagger}_{-\frac32,i} - c^{1\dagger}_{\frac12,i} c^{1\dagger}_{-\frac12,i}  + 
c^{2\dagger}_{\frac32,i} c^{2\dagger}_{-\frac32,i} - c^{2\dagger}_{\frac12,i} c^{2\dagger}_{-\frac12,i}
\\ + c^{1\dagger}_{\frac32,i} c^{2\dagger}_{-\frac32,i} - c^{1\dagger}_{\frac12,i} c^{2\dagger}_{-\frac12,i} + c^{2\dagger}_{\frac32,i} c^{1\dagger}_{-\frac32,i} - c^{2\dagger}_{\frac12,i} c^{1\dagger}_{-\frac12,i}  \big). 
\end{multline}

Similarly, in the bonding-Haldane state, the generators of the hidden SU(2) symmetry, that emerges for the coupling $g_0 = 5 g_2$ for $f_l=3/4$ filling corresponds to the singlet pair and the particle number operator of the bonding quasi-particles:  $\tilde{\mathcal{S}}_i^\dagger = P_{00,i}^{b\dagger}$, and $\tilde{\mathcal{S}}_i^z = (n_i^b - 2)/2$. The bonding singlet pairs do not show quasi-long range order along the ladder, instead the nonzero string order parameter 
\begin{equation}
\label{eq:string2_op}
\mathcal{O}_{HI}^b = \mathrm{lim}_{|i-j|\rightarrow \infty} \left< \tilde{n}_i^b \mathrm{e}^{\mathrm{i} \pi \sum_{i<k<j} \tilde{n}_k^b} \tilde{n}_j^b \right>
\end{equation} 
can identify the state. Here $\tilde{n}_i^b = \tilde{\mathcal{S}}_i^z$.  The edge states in this case are very similar to the ones at quarter filling, the only difference comes from the density profile of the pairs along the two end rungs: in this case the density of the pairs is finite at the center of the rungs.

\subsubsection{General fillings}

At incommensurate fillings in case of fully filled lower band, too, five different phases are expected. A Luttinger liquid  state of the bonding quasi-particles characterizes the system when $g_0\geq g_2 >0$. In the region of $g_0>3g_2$ and $g_2<0$ the Luttinger parameter $K_c$ determines the emerging state. For $K_c < 2$ the ground state is a $4k_F$ density wave with the order parameter \eqref{eq:rung-dw-op}. When $K_c > 2$ the ground state is a molecular superfluid state characterized by SU(4) singlet quartets, similarly that we have seen in the empty upper band case. The MS order parameter now sightly differs from Eq.~\eqref{eq:ms-op}: 
\begin{equation}
\mathcal{O}_\mathrm{quart}= \frac14 \sum_{j_l} c^{j_1\dagger}_{\frac32,i} c^{j_2\dagger}_{\frac12,i} c^{j_3\dagger}_{-\frac12,i} c^{j_4\dagger}_{-\frac32,i}.
\end{equation}
When $g_0 > 3g_2$ and $g_0<g_2$, depending on the value  of $K_c$, two different phases appear. For $K_c > 1/2$ site singlet pairs of the bonding quasi-particles are formed with order parameter $\mathcal{O}_{SF}^b=P_{00,i}^{b\dagger}$. These pairs are coexisting rung and site singlets preserving the translation invariance along the ladder:
\begin{multline}
\label{eq:bcs-b-op}
 \mathcal{O}_{SF}^b=  \frac12  \big( c^{1\dagger}_{\frac32,i} c^{1\dagger}_{-\frac32,i} - c^{1\dagger}_{\frac12,i} c^{1\dagger}_{-\frac12,i}  + 
c^{2\dagger}_{\frac32,i} c^{2\dagger}_{-\frac32,i} - c^{2\dagger}_{\frac12,i} c^{2\dagger}_{-\frac12,i}
\\ + c^{1\dagger}_{\frac32,i} c^{2\dagger}_{-\frac32,i} - c^{1\dagger}_{\frac12,i} c^{2\dagger}_{-\frac12,i} + c^{2\dagger}_{\frac32,i} c^{1\dagger}_{-\frac32,i} - c^{2\dagger}_{\frac12,i} c^{1\dagger}_{-\frac12,i}  \big). 
\end{multline}
Since the pairs consist of two bonding particles, the pair density is large at the center of the rungs, and the pair density profile is like in Fig.~\ref{fig:densities}(a).  For $K_c<1/2$ a (2-atom) molecular density wave of the bonding particles appears with $8k_F$ wave vector. The order parameter $\mathcal{O}_{MDW}^b = e^{\mathrm{i}4k_b R_i} \sum_{\sigma,\sigma'} b_{\sigma,i}^\dagger b_{\sigma',i}^\dagger b_{\sigma',i} b_{\sigma,i}$ in the ladder basis is:
\begin{equation}
\mathcal{O}_{MDW}^b =  e^{\mathrm{i}8k_F R_i} \frac14 \sum_{j_l} \sum_{\sigma,\sigma'}  c^{j_1\dagger}_{\sigma,i} c^{j_2\dagger}_{\sigma',i} c^{j_3}_{\sigma',i} c^{j_4}_{\sigma,i}.
\end{equation} 
This order parameter describes coexisting symmetric and skewed rung centered and site centered molecular density waves of the bonding particles with $8 k_F$ wave vector and without phase shift.

At commensurate fillings the site centered density waves can be suppressed by bond centered ones due to the relevant umklapp processes. At $f_l=2/3$ $4k_F$-BDW order occurs described by the order parameter Eq.~\eqref{eq:plaquette-op} with three-fold spatial periodicity, since $e^{\mathrm{i} 4 k_F R_i}=e^{\mathrm{i}2\pi/3a_0}$. Accordingly, at $f_l=2/3$ filling an oblong plaquette order emerges, similar to the one shown in Fig.~\ref{fig:subfiguresExample3}(a) with 3$a_0$ periodicity. At 5/8-filling, too, instead of the $4k_F$-DW, a $4k_F$-BDW state can emerge. Since $4k_F=4\pi/8 a_0$, now the occurring oblong plaquette order has 4$a_0$ periodicity along the ladder. Note, that all the oblong plaquette states are expected to be insulating preserving the insulating character of their single chain analogous. For $K_c < 1/2$ the $16 k_F$ umklapp processes in the bonding band can become relevant at $f_l=5/8$ filling. In this case instead of the $8k_F$-MDW state, bond centered molecular density wave (BMDW) appears characterized by the order parameter:
\begin{multline}
\mathcal{O}_{BMDW}^b =  e^{\mathrm{i}8k_F R_i} \frac18 \sum_{j_l} \sum_{\sigma,\sigma'}  \\ \cdot (c^{j_1\dagger}_{\sigma,i} c^{j_2\dagger}_{\sigma',i+1} c^{j_3}_{\sigma',i} c^{j_4}_{\sigma,i+1} + c^{j_1\dagger}_{\sigma,i+1} c^{j_2\dagger}_{\sigma',i} c^{j_3}_{\sigma',i+1} c^{j_4}_{\sigma,i}).
\end{multline} 
This order parameter describes two independent $8k_F$-BMDW which leads to a spatially homogeneous molecular valence bond solid state that is similar to the one illustrated in Fig.~\ref{fig:subfig5}.

\section{Conclusions}

In this work we analyzed the possible phases of the relatively strongly coupled spin-3/2 Hubbard ladder. Since the particles in a charge neutral atomic cloud interact mainly via a short range van der Waals interaction the main coupling between the chains is the hopping along the rung of the ladder. We show that when the upper  band of the ladder is empty, or the lower band is full, the system can be described by an effective one-band model. Accordingly, the band degree of freedom is frozen out, and the low energy physics is always determined by the quasi-particles in the anti-bonding (for $f_l<1/2$) or in the bonding (for $f_l>1/2$) state. While in the latter case the density of the quasi-particles along the rungs has a maximum at the sites and zero weight at the center of the rung, in the former case the quasi-particle density has a finite weight at center of the rungs as well. Therefore, despite the similarities of the two cases, different orders characterize them, and depending on the filling parameter and the strength of the on-site interaction in the singlet ($g_0$) and ($g_2$) quintet channel various phases can be stabilized. 

We found that for incommensurate fillings a 4-component Luttinger liquid state is the ground state in an extended region of the parameter space showing that the Luttinger liquid state of the single chain is robust, and stable against large tunneling between two chains. Additionally, for incommensurate fillings depending on the $g_0$, $g_2$ couplings two distinct superfluid states can be the ground state. One of them can be characterized by 2-particle SU(2) singlet Cooper pairs, while the other one is a molecular superfluid state with 4-particle SU(4) singlet quartets. We also found that the charge Haldane insulator state of the one-band spin-3/2 chain survives, too, in the presence of interchain tunneling, and remains the ground state of the ladder when the filling is $f_l=1/4$ or $f_l=3/4$. In the former case the string order and also the free edge states are formed by the particles in the anti-bonding states and in the latter case they consist of particles in the bonding state. Besides, we characterized the various emerging density wave orders. We show that the bonding or antibonding nature of the quasi-particles can lead to serious consequences even on fundamental level, and several inhomogeneous metallic states which usually formed by the antibonding particles can have insulator counterpart formed by the bonding particles.

\section*{Acknowledgements}

We would like to thank Maciej Lewenstein for useful discussions.
This work was supported by the National Research Fund (OTKA) No. K105149 and K100908.

\bibliography{lowdim,ref}

\end{document}